\documentclass[aps,prb,twocolumn,showpacs]{revtex4}
\usepackage{graphicx}
\usepackage{dcolumn}
\newcommand{\bk}{{\bf k}}
\newcommand{\bq}{{\bf q}}

\begin{document}

\title{Spectral Weights, ${d}$-wave Pairing Amplitudes, and Particle-hole 
Tunneling Asymmetry of a Strongly Correlated Superconductor}

\author{Chung-Pin Chou$^{1,2}$, T. K. Lee$^{2}$, and Chang-Ming Ho$^{3}$}
\address{$^{1}$Department of Physics, National Tsinghua University,
Hsinchu 300, Taiwan\\
$^{2}$Institute of Physics, Academia Sinica, Nankang 115, Taiwan\\
$^{3}$Department of Physics, Tamkang University, Taipei 251, Taiwan}

\date{\today}

\begin{abstract}
The spectral weights (SW's) for adding and removing an electron of the Gutzwiller projected 
$d$-wave superconducting (SC) state of the $t$-$J$-type models are studied 
numerically on finite lattices. Restrict to the uniform system but treat exactly the 
strong correlation between electrons, we show that the product of weights is equal to the 
pairing amplitude squared, same as in the weakly coupled case. In addition, we derive 
a rigorous relation of SW with doping in the electron doped system and obtain 
particle-hole asymmetry of the conductance-proportional quantity within the SC gap energy and, also, the 
anti-correlation between gap sizes and peak heights observed in tunneling spectroscopy on 
high $T_c$ cuprates.
\end{abstract}

\maketitle

The emergence of the superconductivity as holes doped into the Mott insulating parent 
compounds is one of the intriguing phenomena of high $T_c$ cuprates \cite{bonn2006}. 
It is usually emphasized, however, below the transition temperature there is strong 
similarity of the superconducting (SC) state with that of low $T_c$ materials in 
the sense that the state is composed of Cooper pairs of electrons, though with 
unconventional $d$-wave pairing symmetry \cite{tsuei-kirtley} and thin superfluid density 
\cite{gossamer}. Nevertheless, differences from the conventional SC state are unearthed 
clearly by high-resolution scanning tunneling microscopy and spectroscopy (STM/STS) on 
different cuprates with easily-cleaved surfaces. Namely, despite of physical quantities 
with nanometer scale inhomogeniety, the averaged (over some area within the scanned 
field-of-view) tunneling conductance is of unexpected behavior: its amplitude at negative 
sample bias- {\it for removing electrons} - is often larger than that at positive one- {\it for 
adding electrons} - \cite{stm0,stm1,fang,stm2}. Most intriguingly, the gap size, which 
was usually inferred from photoemission experiments \cite{arpes}, can be 
extracted directly from conductance peaks for the first time and is found to be larger as 
peaks become less pronounced \cite{stm1,fang,stm2}. Same behaviors have been observed 
previously in underdoped cuprates using point-contact tunneling set-up \cite{pct}.

The theoretical attempt to understand the features seen in STM/STS {\it hitherto} is 
mostly about the effects of the inhomogeneous dopant induced SC order parameters to the 
local density of states (LDOS) of the $d$-wave BCS ($d$-BCS) state \cite{hirschfeld-su},
it largely neglects the strong correlation between electrons which should be essential for the 
case of underdoped cuprates \cite{lee-nagaosa-wen}. Actually, there is 
not enough understanding about the effect of strong correlation for a homogeneous system to help us 
 to address the complex issues of disorder as revealed by  
tunneling experiments. Recently, Anderson proposed that the asymmetric tunneling conductance is 
closely related to the strong correlations inherent in the 
Gutzwiller projected $d$-BCS or, simply, resonating-valence-bond 
wave function (RVB WF) \cite{pwa-ong}. 
However, treating the projection only approximately by the usual scheme of the renormalized 
mean-field theory (RMFT), there have been controversy whether the asymmetry is accounted 
for by the coherent quasi-hole (QH) and -particle (QP) excitations of the projected state or rather by 
the incoherent part dictated by the spectral sum rule \cite{rantner-wen,sum-rule,fl}. Also, the 
correlation between gap sizes and peak heights has not yet been examined clearly from the 
strong correlation point of view \cite{wang-han-lee-other}. 

In this paper, we defer the issue of inhomogeneity to later work and 
examine exactly the effects of strong correlation by numerically investigating the spectral 
weights (SW's) of the $d$-wave RVB ($d$-RVB) state on finite square lattices.  
With SW, particularly, $Z_{\bf{k}\sigma}^{-}$ for removing an electron (defined in 
Eq.(\ref{e:sw1}) below), calculated, we obtained several results:  
(i) $d$-wave pairing amplitude squared is equal to the products of SW's, as it is exactly 
for weakly-coupled case; (ii) inspired by the hole doped case we focus mostly here, a rigorous 
relation of SW for removing an electron is derived for the electron doped case; 
(iii) the difference between $Z_{\bf{k}\sigma}^{-}$ calculated exactly and by using RMFT becomes 
significant at low, finite doping of holes;  
(iv) the particle-hole asymmetry of the sum of low-energy SW's over momenta within 
some energy window near the Fermi level becomes more prominent with reduced doping \cite{FLEX} and,  
meanwhile, (v) the heights of SW peaks decrease as the gap sizes increase.

Let us start by defining the SW for adding (and removing) one electron we 
calculate, {\it i.e.}   
\begin{eqnarray}
Z_{\bk\sigma}^{+(-)} = \frac{\mid\langle N_e{+(-)}1\mid
c^{\dagger}_{\bk\sigma}(c_{\bk\sigma})\mid N_e\rangle\mid^2}{\langle N_e\mid
N_e\rangle\langle N_e{+(-)}1\mid N_e{+(-)}1\rangle}, \label{e:sw1}
\end{eqnarray}
where, for momentum {\bf k},  
\begin{equation}
\mid N_e+1\rangle \equiv P_dc^{\dagger}_{\bk\sigma}\mid
{N_e}\rangle_{0}
\end{equation}
for the QP excitation, and 
\begin{equation}
\mid N_e-1\rangle \equiv P_dc^{\dagger}_{-\bk-\sigma}\mid
{N_e-2}\rangle_{0}
\label{e:h-exc}
\end{equation}
for the QH one which is also proportional to 
$P_dc_{\bk\sigma}\mid N_e\rangle_{0}$. Here 
$\mid N_e\rangle_{0}$ is related to the trial WF of the projected 
electron-paired ground state in a uniform system, 
\begin{eqnarray}
&\mid N_e\rangle&=P_d\mid N_e\rangle_{0} \equiv P_d\left(\sum_{\bq}a_\bq
c^\dagger_{\bq\uparrow}c^\dagger_{-\bq\downarrow}\right)^{N_e/2}\mid
0\rangle .\label{e:twf}
\end{eqnarray}
The variationally optimized 
$\mid N_e\rangle$ we focus on in this paper is the $d$-RVB state \cite{vmc}. 
With $N_e$ the total 
number of electrons, coefficient $a_\bq=v_{\bq}/u_{\bq}=(E_\bq-\epsilon_\bq)/\Delta_\bq$ in 
which $v_{\bq}$ and $u_{\bq}$ are SC coherent factors, $\epsilon_{\bq}=-(\cos q_x+\cos q_y)-
\mu-t_{\textit{v}}'\cos q_x\cos q_y-t_{\textit{v}}''(\cos 2q_x+\cos 2q_y)$, 
$\Delta_\bq=\Delta_{v}(\cos q_x-\cos q_y)$, and
$E_\bq=\sqrt{\epsilon_\bq{^2} + \Delta^2_\bq}$. The operator $P_d$ projects out 
the doubly-occupied sites in the system with a finite number of doped holes present. In addition 
to $\Delta_{v}$ and $\mu$, $t_{\textit{v}}'$ and $t_{\textit{v}}''$ are the other two variational 
parameters associated with the long-range hoppings in  the $t$-$t'$-$t''$-$J$ model Hamiltonian,  
$H=-\sum_{ij}P_{d}t_{ij}({c}^\dagger_{i,\sigma}{c}_{j,\sigma}
+ h.c.)P_{d}+J\sum_{<i,j>}({\vec{S}_i\cdot\vec{S}_j}-\frac{1}{4}n_{i}n_{j})$, 
where hopping amplitude $t_{ij}$=$t$, $t'$, and $t''$ for sites $i$ and $j$ being the nearest-,
next-nearest-, and the third-nearest-neighbors, respectively, $\vec{S}_{i}$ the spin operator 
at site $i$, $<i,j>$ means that the interaction between spins occurs only for the 
nearest-neighboring sites. 

Applying identities for projection operator, 
\begin{eqnarray}
[c_{\bk\sigma},P_d]P_d &=& 0 ;\nonumber\\
P_d c_{\bk\sigma}[c^{\dagger}_{\bk'\sigma},P_d]&=&
P_d[\frac{1}{N}\sum_{i}e^{i({\bk}'-{\bk})\cdot\vec{R}_
{i\sigma}}n_{i,-\sigma}]P_d
\label{e:identity1}
\end{eqnarray}
with $\vec{R}_{i\sigma}$ the position vector of the $i$-th spin $\sigma$ in 
the lattice of size $N$ and $n_{i\sigma}=c^{\dagger}_{i\sigma}c_{i\sigma}$,
we can relate the $Z_{\bk\sigma}^{+}$ exactly to the momentum
distribution function (MDF) $n_{\bk\sigma}$ as
\begin{equation}
Z_{\bk\sigma}^{+}=\frac{1+x}{2}-n_{\bk\sigma}, \label{e:relation}
\end{equation}
where {\it x} is the density of doped holes and $n_{\bf k}=\langle N_e\mid
c^{\dagger}_{\bk\sigma}c_{\bk\sigma}\mid
N_e\rangle / \langle N_e\mid N_e\rangle$ \cite{yunoki,nave,yang}. 

As a digression to electron doped (ED) case, it is straightforward to show that, 
applying the hole-particle transformation to Eq.(\ref{e:relation}) \cite{e-doped}, 
SW of removing an electron in ED system satisfies rigorously 
$Z^{-}_{\bk\sigma}$=$n_{\bk\sigma}-(1-x)/2$. This relation may be verified in 
experiment.  

\begin{figure}[top]
\rotatebox{0}{\includegraphics[height=2.0in,width=3.0in]{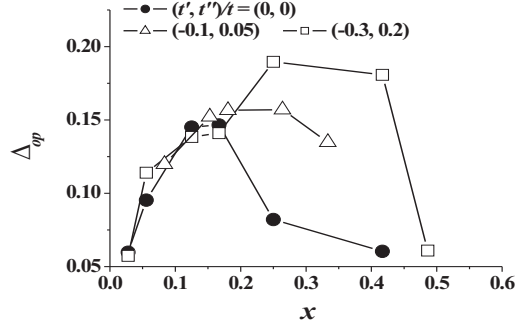}}
\caption{The SC pairing amplitude for $d$-RVB state as a function of doping determined 
by the products of SW's using 
Eqs.(\ref{e:pairing}) and (\ref{e:op}). System size here is $12\times12$. 
Different symbols represent results obtained for different values of $(t',t'')/t$, 
as indicated} 
\label{f:Fig1}
\end{figure}

Back to the hole doped case, although there is no exact relation like this for $Z_{\bk\sigma}^{-}$, 
we notice  that $Z_{\bk\sigma}^{+}$ and $Z_{\bk\sigma}^{-}$ satisfy a relation, 
\begin{equation}
Z_{-\bk-\sigma}^{+} \cdot Z_{\bk\sigma}^{-} =\frac{\mid\langle N_e\mid
c^{\dagger}_{\bk\sigma}c^{\dagger}_{-\bk-\sigma}\mid
N_e-2\rangle\mid^2}{\langle N_e\mid N_e\rangle\langle N_e-2\mid
N_e-2\rangle} \equiv P_{\bk}
\label{e:pairing}
\end{equation}
which can be proved straightforwardly by combining Eqs.(\ref{e:h-exc}) and
(\ref{e:identity1}). The matrix element $P_{\bk}$, which represents the 
off-diagonal long-range order in the pairing correlation, 
is related to the $d$-wave SC pairing 
amplitude or order parameter $\Delta_{op}$ by  
\begin{equation}
\Delta_{op}=\frac{2}{N}\sum_{\bk}\mid
\cos k_x-\cos k_y\mid\sqrt{P_{\bk}}. \label{e:op}
\end{equation}
With both the SW's computed numerically, we plot in Fig.\ref{f:Fig1} 
the doping dependence of $\Delta_{op}$ which indeed has the {\it dome}-like shape 
similar to the $T_{c}$ versus doping determined experimentally. Actually, the peak 
positions shown in Fig.\ref{f:Fig1} are almost the same as what have been 
obtained previously by studying the $d$-wave long-range pair-pair correlation \cite{vmc,overestimate}. 
Furthermore, with more holes doped into the system, just like the reduction of long-range 
correlation between electron pairs is induced by the change of the anti-nodal Fermi surface 
geometry \cite{vmc}, the SC order parameter is decreased due to $Z_{\bk\sigma}^{-}$ with {\bf k} 
near $(\pi,0)$. Hence Eq.(\ref{e:pairing}) provides another way to evaluate the strength of 
the pairing amplitude.

For the BCS theory without projection, we know 
$Z_{\bf{k}\sigma}^{+(-)}=u_{\bf k}^{2}(v_{\bf k}^{2})$ and Eq.(\ref{e:pairing}) is also 
exactly satisfied. For the strongly correlated $t$-$J$-type models, even though 
the same relation is followed in RMFT \cite{rmft}, it is still surprising 
to find out that this relation is correct in the RVB state with projection rigorously obeyed. 

On the other hand, reminiscent of what have been argued previously by analytic approach \cite{rantner-wen}, 
we recognize that the strong correlation 
effects becomes apparent only in $Z_{\bk\sigma}^{-}$ at low doping.
The effects due to strong correlation are 
examined by comparing the {\it coherent} SW's averaged over all momenta, {\it i.e.} 
$Z^{-}_{ave}\equiv\sum_{\bk}Z_{\bk\sigma}^{-}/N$, and the {\it incoherent} part defined by the relation 
\begin{equation}
n_{ave}^{inc}\equiv n_{ave}-Z^{-}_{ave}  
\label{e:incoherent}
\end{equation}
obtained by exact treatment of the projection and by using RMFT. Here 
$n_{ave}\equiv\sum_{\bk}n_{\bk\sigma}/N$ is the average MDF which should always be equal to the 
electron density of the system.  

\begin{figure}[top]
\rotatebox{0}{\includegraphics[height=2.0in,width=2.6in]{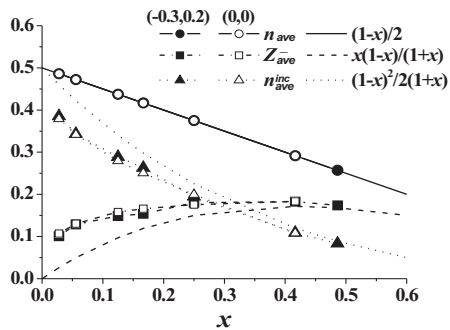}}
\caption{The doping dependence of SW's for removing an electron and average MDF 
of $d$-RVB state obtained numerically for  
the $12\times12$ lattice and by RMFT. Numerical and expected $n_{ave}$ are 
denoted by the circles and the solid line, respectively. 
The squares (triangles), connected by dashed (dotted) line as the guide for the eye, 
are for exact results of $Z^{-}_{ave}$ [$n_{ave}^{inc}$, extracted using 
Eq.(\ref{e:incoherent}) with $n_{ave}$ calculated numerically]. All solid symbols are 
results for $(t',t'')/t$=(-0.3,0.2) and empty ones for (0,0). The dashed and dotted 
lines without data points represent results by RMFT.}
\label{f:Fig2}
\end{figure}

The exact results for the $12\times12$ lattice and that by RMFT are shown in Fig.\ref{f:Fig2}. 
The coherent part of $Z_{\bk\sigma}^{-}$ by RMFT is 
$g_{t}v^{2}_{\bk}$ with renormalization factor $g_{t}=2x/(1+x)$. 
Completing the momentum sum for the coherence factor, the average result is $x(1-x)/(1+x)$ and, thus, 
$n_{ave}^{inc}=[(1-x)^{2}]/2(1+x)$ \cite{sum-rule}, plotted in Fig.\ref{f:Fig2} (dashed 
and dotted lines, respectively) in comparison with the exact ones. As is shown there, while 
the numerical $n_{ave}$ (solid circles) is indeed equal to the electron density, the exact incoherent 
SW for removing an electron is less than the RMFT result. The difference becomes more 
significant as 
hole doping level is reduced. Interestingly, this behavior is independent of the $(t',t'')/t$ 
values (represented by solid and empty symbols in Fig.\ref{f:Fig2}) which correspond to very 
different doping dependence of the Fermi surface shape 
and also the DOS. By contrast, the average values of $Z_{\bk\sigma}^{+}$ calculated exactly (not 
shown) and by RMFT are identical due to Eq.(\ref{e:relation}). 
 
\begin{figure}[top]
\rotatebox{0}{\includegraphics[height=2.0in,width=3.0in]{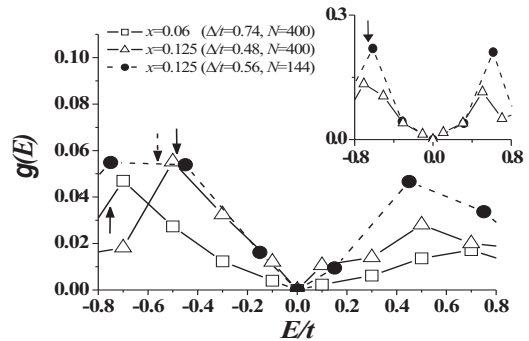}} 
\caption{$g(E)$ for $d$-RVB state versus excitation energy $E$ for $12\times12$ (solid circle) 
and $20\times20$ lattices (empty square and triangle, for different dopings). 
The associated excitation gap positions are marked by arrows (see text). 
\textbf{Inset}: Same plots for $d$-BCS state with gap value 0.6 and $x=0.125$. 
The data shown here are 
for different sizes, denoted by the same symbols as in the main figure, with parameters 
$\mu/t=-0.269$ ($12\times12$) and $\mu/t=-0.272$ ($20\times20$), respectively. 
All data are obtained for $(t',t'')/t$=(-0.3,0.2). }
\label{f:Fig3}
\end{figure}

To make a comparison with tunneling experiments, we then concentrate on the 
SW's as a function of the excited-state energy. By applying the model 
Hamiltonian to excitations $\mid N_e\pm 1\rangle$, we calculate their 
excitation energies for each momentum and also the corresponding energy 
gap by fitting the excitation energy $E_{\bf k}$. To reduce the effect of finite size, we define the 
sum of $Z_{\bk\sigma}^{\pm}/N$, over momentum {\bf k} which has energy within $E-{\Delta E}/2$ and 
$E+{\Delta E}/2$, as $g(E)$ [negative(positive) for removing(adding) an electron] which could be 
viewed, approximately, as proportional to the conductance at low energy $E$. We plot 
$g(E)$ in Fig.\ref{f:Fig3}, up to about the energy where peaks appear for lattices of size $12\times12$ with 
$\Delta E/t=$0.3, and also $20\times20$ with energy interval 0.2 for various dopings \cite{e-interval}. 
To make sure our treatment is correct, we have also applied the same analysis to 
the {\it d}-BCS state. As shown in the inset of Fig.\ref{f:Fig3}, the ideal BCS result is 
hardly distorted by the finite size. 
Note that, with the reasonable finite-size dependence, we obtain indeed the {\it V}-shape 
$d$-wave gap near zero energy. The width between peak positions is also roughly equal to 
two times of the gap value deduced from the excitation energy. Looking at the results closely, while $g(E)$ 
may indeed be about the same at the opposite sides in the very vicinity of zero energy as 
suggested in Ref.22, $g(E)$ for removing an electron is always larger than that for adding an electron 
at higher energy near that of the peak. With decreased doping, the ratio of $g(E)$ at negative and positive energies 
enhances quite dramatically, {\it e.g.} from $x$=0.125 to 0.056, 
$g(-{\Delta})/g({\Delta})$ at the corresponding energy of the peak $\Delta$ (in units of $t$)
increases from 1.96 to 2.73 \cite{overestimate}. Similar behaviors are found for the case with vanishing 
$(t',t'')/t$ (not shown). In contrast to this, for the $d$-BCS (inset in Fig.\ref{f:Fig3}) 
case in the same finite lattices there is almost no change of the ratio 
within the gap. The numerical results thus tells us features due to strong correlation which 
are not fully explored yet in the tunneling experiment, {\it i.e.} 
the particle-hole asymmetry of average conductance exists even within the gap region and gets 
enhanced with underdoping.    


\begin{figure}[top]
\rotatebox{0}{\includegraphics[height=2.0in,width=2.8in]{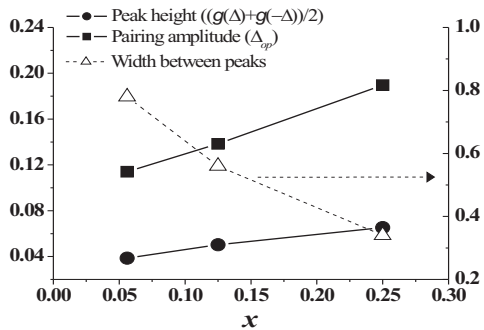}} 
\caption{The doping dependence of the peak height for $g(E)$, pairing 
amplitude and width between peaks in $20\times20$ lattice for $d$-RVB case. Data 
are extracted from that shown in Fig.\ref{f:Fig3} and those obtained in the same way but 
not shown there. The values for the width between 
peaks are referred to the vertical axis on the right-hand side (indicated by the arrow). 
}
\label{f:Fig4}
\end{figure}

Fig.\ref{f:Fig3} also reveals correlation between 
heights of the spectral weight peak and the gap size (or the width between peaks) as doping level 
is varied \cite{overestimate}. Within the doping level shown in Fig.\ref{f:Fig4}, the peak height 
scales with the pairing amplitude
but apparently anti-correlates with the gap size. This is in 
clear contrast to the BCS case in which the peak height, proportional to the 
SC coherence, scales with the width between peaks or gap size as more holes doped 
into the system. Our result agrees qualitatively with what has been extracted 
from STS experiments \cite{stm1}. 

To conclude, in order to provide a better understanding of the results measured by the tunneling 
experiments without the complication of mixing disorder and strong correlation, 
here we studied the SW's for adding and removing an electron for a uniform $d$-RVB SC 
state without disorder. We derive analytically and examine numerically the relation 
between pairing amplitude and SW products. Performing particle-hole transformation, we obtain also 
exact dependence of SW for removing an electron with doping in the ED systems, which could be tested 
by photoemission spectroscopy. While the strong 
correlation effect is less noticeable by looking at the pairing amplitude, we found that 
the SW for removing an electron deviates clearly from results obtained by RMFT 
in the low doping regime. More specifically, at this doping level the conductance-related quantity 
of the uniform $d$-RVB state on finite lattices computed exactly is particle-hole asymmetric below 
the gap energy and, consistent qualitatively with what is seen in recent tunneling 
experiments, the extracted gap value (from the excitation energy) or, equivalently, the width 
between SW peaks anti-correlates with the peak heights.

We acknowledge discussions with Profs. M. Ogata, N. Nagaosa, T. Tohyama, 
S. Ishihara, V.N. Muthukumar and Drs. Y. Yanase and N. Fukushima. CMH is 
grateful to Profs. S.~H. Pan, H. Takagi and Dr. T. Hanaguri for kindly sharing their 
results and insights about STM/STS. TKL and CMH are supported by the National 
Science Council in Taiwan with Grant no.94-2112-M-001-003 and 94-2112-M-032-001, 
respectively. Part of the calculations are performed in the IBM P690 and SMP2 in 
the Nation Center for High-performance Computing in Taiwan.



\begin{references}
\bibitem{e-interval}The energy interval we chose is to reduce the effects due to 
lattice size and also to include about the same number of momentum points below and above zero 
energy as we calculate the average weights of adding and removing an electron, respectively.
\end{references}
\end{document}